\documentclass[conference]{IEEEtran}
\usepackage{nopageno}
\usepackage[T1]{fontenc}
\usepackage{fourier}
\usepackage{float}
\usepackage[english]{babel}															
\usepackage[protrusion=true,expansion=true]{microtype}	
\usepackage{amsmath,amsfonts,amsthm} 
\usepackage[pdftex]{graphicx}	
\usepackage{url}
\usepackage{syntax}
\usepackage{array}
\usepackage{txfonts}
\usepackage{listings}

\usepackage{fancyhdr}
\numberwithin{equation}{section}		
\numberwithin{figure}{section}			
\numberwithin{table}{section}				

\title{
        Data-driven Workflows for Microservices
}

\author{
\IEEEauthorblockN{Larisa Safina\IEEEauthorrefmark{1}, Manuel Mazzara\IEEEauthorrefmark{1}, Fabrizio Montesi\IEEEauthorrefmark{2}}
    \IEEEauthorblockA{\IEEEauthorrefmark{1}Innopolis University, Russia
    \\\{l.safina, m.mazzara\}@innopolis.ru}
    \IEEEauthorblockA{\IEEEauthorrefmark{2}University of Southern Denmark
    \\ fmontesi@imada.sdu.dk }
}

\begin{document}
\maketitle
\begin{abstract}

Microservices is an architectural style inspired by service-oriented computing that has recently started gaining popularity. Jolie is a programming language based on the microservices paradigm: the main building block of Jolie systems are services, in contrast to, e.g., functions or objects. The primitives offered by the Jolie language elicit many of the recurring patterns found in microservices, like load balancers and structured processes. However, Jolie still lacks some useful constructs for dealing with message types and data manipulation that are present in service-oriented computing. In this paper, we focus on the possibility of expressing choices at the level of data types, a feature well represented in standards for Web Services, e.g., WSDL. We extend Jolie to support such type choices and show the impact of our implementation on some of the typical scenarios found in microservice systems. This shows how computation can move from a process-driven to a data-driven approach, and leads to the preliminary identification of recurring communication patterns that can be shaped as design patterns.


\end{abstract}
\section{Introduction}
The increasing complexity of modern software requires new approaches to architectural design and system modeling. Complex systems also show high level of concurrency, i.e. multiple intertwined threads of executions, often running on different hardware, which need to be synchronized and coordinated, and which much share information, often through different paradigms of communication. Therefore improving software quality and deploying reliable services is the consequence of an accurate use of optimal service-based architectural styles and well-established software engineering techniques for requirements elicitation, design, testing and verification, in particular when it come to concurrent service-based systems.
In order to tackle these issues from the architectural viewpoint, Microservices architecture appeared lately as a new paradigm for programming applications by means of the composition of small services, each running its own process and communicating via light-weighted mechanisms \cite{fowler}. This approach has been built on the concepts of Service-oriented Architectures\cite{soa} brought from crossing-boundaries workflows to the application level, and into the applications architectures, i.e. it Service-oriented Architecture and Programming from the large to the small.

Microservices architecture still shows distinctive characteristics which blend into something unique and different from SOA itself. The size is comparatively small versus a typical service \cite{fowler}, supporting the belief that 
the architectural design of a system is highly dependent on the structural design of the organization producing it.

In the context of microservices the Jolie programming language \cite{jolie,jolie:website} emerged as a paradigmatic solution tuned at getting the best out of this architectural style. Jolie is comprehensive and capable of implementing both simple services and complex orchestrations. Everything in Jolie is a microservice and all these microservices can be easily reused or composed for obtaining in turn new microservices. This approach supports distributed architecture and guarantees simple managing of components, which reduces maintenance and development costs.

This work is devoted to extend the Jolie programming language in order to support data-driven workflow, i.e. moving the control-flow decision-making from process-driven to data-driven. That means that control-flow can be directed at the time of message passing according to the nature of the message strucutre and type, instead of requiring post-reception processing. A typical example of process-driven workflow is presented in \cite{Mazzara11}. In the microservice scenario this opportunity opens to novel programming patterns. Whether the process-driven or data-driven programming style is more suitable for an application really depends on the specific problem domain and, to same extent, to developers preferences and style. Data-driven flows have been realized in Jolie by means of an extension to its type system, which manifests as implementation of choice type. Further extensions can be implemented in order to strengthen further this possibility.

Two major contributions appear as a result of this work. The first is of scientific nature, i.e., moving from process-driven to data-driven and therefore open to new programming patterns and styles. The second is of purely technical interest, and stands in the re-engineering of the Jolie interpreter as a consequence of the extended data type. This represents a relevant case study of interpreter re-engineering, and therefore can be valuable experience for the practitioners involved in activities of comparable complexity.

The paper is structured as follows: in section \ref{sec:jl} a quick overview of Jolie programming language is given in order to intorduce the unfamiliar reader with the main concepts and syntax. Far from being an exhaustive report, the section is just a compendium providing the links for further study of the topic. Section \ref{sec:case} presents a case study on top of which the major narrative of the paper is built and the contributions are defined. In particular, two different kind of approaches computation are described: process-driven and data-driven, and examples of both are explained in order to understand the differences. Sections \ref{sec:interpreter} and \ref{sec:data-driven} describe the architecture of the Jolie interpreter and the changes that were necessary in order to extend the type system. Finally, section \ref{sec:conclusions} wraps up final considerations regarding the contribution of this work, and presents ideas on how future developments may be built up on top of current achievements.

\section{Jolie language}
\label{sec:jl}
In this section we brielfy recall the Jolie programming language in order to simplify the reading of the remaining part of the paper.

Each Jolie program is composed by two parts, a behavioral part and a deployment part. A program can be formally expressed as: 

\vspace{0.2cm}
{\setlength\tabcolsep{4pt}
\begin{tabular}{l >{$}c<{$} c c}
  Program &\Coloneqq & Deployment & Behavior\\
\end{tabular}}

\vspace{0.2cm}
\subsubsection{Deployment part}
The deployment part contains directives which help the Jolie program to receive and send messages and be orchestrated among other microservices. The deployment part is separated from the program behavioral part, so that the same behavior may be reused later with a different deployment configuration. Formally the deployment part is expressed as:

\vspace{0.2cm}
{\setlength\tabcolsep{4pt}
\begin{tabular}{l >{$}l<{$} c}
Deployment &\Coloneqq & DeploymentInstruction*\\
\end{tabular}}

\vspace{0.2cm}
Where \textit{DeploymentInstruction} can include 
\begin{itemize}
\item \textit{Interfaces}: sets of operations equipped with information about their request (and sometimes response) types;
\item\textit{Message types}: can be represented as native types, linked types or undefined. Message types are the main subject of this report and shall be discussed later in section~\ref{sec:jtp}.
\item \textit{Communication ports}: define how communications with other services are actually performed.
\end{itemize}

\vspace{0.2cm}
\subsubsection{Behavioral part}
The behavioral part contains microservice implementation of the functionalities, containing both computations and communication expressions. Examples of expanded behavioral part utilization will be provided later in section~\ref{sec:ex}. 
Formally, behavioral part can be expressed as:

\vspace{0.2cm}
{\setlength\tabcolsep{4pt}
\begin{tabular}{l >{$}l<{$} l}
Behavior &\Coloneqq & BehaviouralBlock* \\
 &  & main $\{$Process$\}$ BehaviouralBlock* \\
BehaviouralBlock &\Coloneqq & define id $\{$Process$\}$ \\
        & | & init $\{$Process$\}$ \\
\end{tabular}}

\vspace{0.2cm}
\textit{Main} is a procedure which defines an entry point of execution. \textit{Main} can be followed or preceded by define procedures with \textit{id} identifier. 

\textit{Init} supports special procedures for initializing a service before it makes its behaviors available. Procedures  specified with \textit{define} can be used many times, while the one specified with \textit{init} is executed only once, when the service is started.

\textit{Process} defines the activities to be performed by the service. Processes can be composed in sequences, parallels and (input guarded) non-deterministic choices \cite{jolie}.

\vspace{0.2cm}
\subsubsection{Communication}
Communication of processes can be performed by two possible patterns: \textit{one-way} (the endpoint receives a message) and \textit{request-response} (the endpoint receives a message, and sends a response back to the caller):
\vspace{0.2cm}

{\setlength\tabcolsep{4pt}
\begin{tabular}{l >{$}l<{$} l l}
Process &\Coloneqq & $\ldots$ | InputStatement \\
& | & OutputStatement  \\
InputStatement &\Coloneqq & op(x) & (One-Way) \\
     & | & op(x)(y) { Process } & (R-Response) \\
OutputStatement &\Coloneqq & op@OPort(x) &(Notification) \\
     & | & op@OPort(x)(y) & (S-Response) 
\end{tabular}}
\vspace{0.2cm}

\textit{One-Way} is used to receive a message for operation \textit{op} in variable \textit{x}.
\textit{Request-Response} is used to receive a message for operation \textit{op} in variable \textit{x}, execute a \textit{Process} and then send back a response to the caller containing the value of variable \textit{y}. 
\textit{Notification} and \textit{Solicit-Response} are the dual of the former ones to be used, respectively, for sending a message to a \textit{One-Way} statement or to a \textit{Request-Response} one. 
\textit{OPort} defines an output port name of desired endpoint. 

\subsection{Jolie type system}
\label{sec:jtp}
Jolie provides a language for describing the types that are allowed to be communicated over a network. Communications are type checked at run-time when a message is received \cite{nielsen}. Message types are introduced in the deployment part of Jolie programs: 

\vspace{0.2cm}
{\setlength\tabcolsep{4pt}
\begin{tabular}{c >{$}l<{$} l}
DeploymentInstruction &\Coloneqq & $\ldots$ \\
    & | & type id : TypeDefinition 
\end{tabular}}

\vspace{0.2cm}
Where \textit{id} is an identifier in order to use the message type in other program parts and \textit{TypeDefinition} can be a native type, native type with subtypes, native type with undefined subnodes, link type or can be undefined (means that variable is null until a value is assigned to it):

\vspace{0.2cm}
{\setlength\tabcolsep{4pt}
\begin{tabular}{c >{$}l<{$} l l}
TypeDefinition &\Coloneqq & $\ldots$ \\
   & | & NativeType \\
   & | & NativeType $\{$ SubTypeList $\}$ \\
   & | & NativeType $\{$ ? $\}$ \\
   & | & id 		\\
   & | & undefined 	 \\
NativeType &\Coloneqq & int | double | string | raw | void | any 
\end{tabular}}

\vspace{0.2cm}
Where $\{ ? \}$ represents untyped subnodes and \textit{undefined} stands a shortcut for any:$\{?\}$.

\vspace{0.2cm}

Except commonly-used native types as int, double or string, Jolie also has the following types:
\begin{itemize}
\item \textit{raw} (used for transmission of raw data streams as byte arrays)
\item \textit{void} (is used for indicating that no value is contained by the variable)
\item \textit{any} (means that any native type with which variable is initialized will be accepted. 
\end{itemize}

Untyped subnodes, expressed as $\{$?$\}$  construction, indicate that a node may have any kind of subtree. Already defined types can be reused in other types definition as link types by means of their ids. 

Type may have any number of subtypes, which bnf-form is the following:

\vspace{0.2cm}
{\setlength\tabcolsep{4pt}
\begin{tabular}{c >{$}l<{$} l l}
SubTypeList &\Coloneqq & SubType \\
 & | & SubType SubTypeList \\
SubType &\Coloneqq & .id Cardinality : TypeDefinition
\end{tabular}}

\vspace{0.2cm}
Each subnode has its cardinality defined as one by default or as following:

\vspace{0.2cm}
{\setlength\tabcolsep{4pt}
\begin{tabular}{c >{$}l<{$} l l}
Cardinality &\Coloneqq & [ int , int ] &(Range) \\
& | & [ int , * ] &(Lower-bound) \\
& | & * & (Shortcut for [0,*]) \\
& | & ? & (Shortcut for [0,1]) \\
& | & $\epsilon$ 
\end{tabular}}

\vspace{0.2cm}
\section{Case Study}
\label{sec:case}
In this section we show how the extension to the Jolie type system led to an enhanced arsenal at developers' disposal. We will proceed by examples. The short compendium of the Jolie syntax as presented in the previous section, combined with the code examples presented here, may be sufficient to grasp a general understanding. For a more comprehensive information the reader can refer to \cite{jolie:website} and \cite{Guidi07}. 

Let us consider an example implementing a car rental. This consists of three parts: 
\begin{itemize}
\item \textit{Server}, which provides the rental service; 
\item \textit{Client}, which wants to use it; 
\item \textit{Interface}, which declares the operations by means of which client and server can interact with each other.  
\end{itemize}

\vspace{0.2cm}
\begin{figure}[H]
\centerline{\includegraphics[scale=0.8]{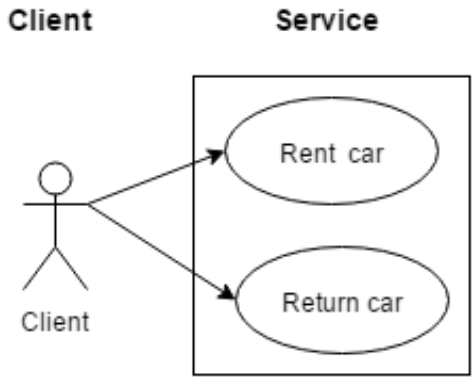}}
\caption{Client-service use-case diagram}
\end{figure}



We wil consider two possible approaches for the server to handle clients' requests: process-driven, implemented in Jolie language by means of input-guarded operations, and data-driven implemented by the newly added choice operator. We will then add some considerations on the usage of both. 

\subsection{Process-driven approach}
In this example, the Interface contains the definition of two operations, dedicated to renting and returning the car, and several data types: 
\begin{itemize}
\item \textit{customer}: stores customers personal information (name, age, and driving license number) necessary to rent a car;
\item \textit{car_return}: holds the reference to customer's profile who has rent this car, the car's identification number and the state of the car after renting.
\end{itemize}

\begin{lstlisting}[basicstyle=\small]
//Car rent interface
type customer: void {
	.name: string
	.age: int
	.license: string
}

type car_return: void {
	.car_state: string
	.c?: customer
	.car_id: string
}

interface CarRentInterface { 
	RequestResponse: 
      get_car( customer )( string )
	RequestResponse: 
      return_car( car_return )( string )
}
\end{lstlisting}

the server indicates the useage of operations defined in the above interface. This is done by the \textit{include} directive at the beginning of the source code. The server's deployment part contains declaration of input port, by means of which it can be accessed with the name and the protocol provided on the defined location. The behavior part of the program contains definition of two operations, \textit{get_car} and \textit{return_car}. They are placed inside the square brackets "[ .. ]", which are used in input-guarded choice syntax. This means, that only one of the operations can be executed at the time, while the others will be deactivated.

\begin{lstlisting}[basicstyle=\small]
//Server with input-guarded operations
include "carRentInterface.iol"

inputPort RentService {
	Location: "socket://localhost:2001"
	Protocol: sodep
	Interfaces: CarRentInterface
}

execution{ concurrent }

main{
	[get_car( request )( response ){	
		response = "43535"
	}]

	[return_car( request )( response ){
	  if (request.car_state == "damaged"){
	    response = "Car is damaged!"
	  } else {
	    response = "Thank you!"
	  }
	}]
}
\end{lstlisting}

The deployment part of the client describes how the connection to the Rent Service works: the same interface, protocol and Rent Service location. The behavioral part of the client program executes the following operations:
\begin{itemize}
\item creating the request information
\item sending this information to Rent Service to be processed by get_car procedure
\item Checking the response and printing it out
\end{itemize}

\begin{lstlisting}[basicstyle=\small]
//Client.ol
include "carRentInterface.iol"
include "console.iol"

outputPort RentService {
	Location: "socket://localhost:2001"
	Protocol: sodep
	Interfaces: CarRentInterface
}

main{
  //sending request for a car
  customer.name = "John Smith";
  customer.age = 32;
  customer.license = "l23454675";

  get_car@RentService( customer )( response );
  println@Console
    ( "Car rent request is accepted. ")();
  println@Console 
    ( "Car id is " + response )()
	
  //returing the car
  return.car_id = response;
  return.car_state = "damaged";
  return_car@RentService( return )( response );
  println@Console
    ( "Car is returned. " + response )()
}
\end{lstlisting}
In order to better understand the execution, we show here the results of running the application:
\begin{lstlisting}[basicstyle=\small]
Process-driven approach
Car rent request is accepted. Car id is 43535
Car is returned. Car is damaged!
\end{lstlisting}

The process-driven approach shows how, classically, the action flows is directed via the use of input-guarded choice. This approach is heavily influenced by the heritage of process algebra, as described in \cite {Guidi07}. Input-guarded choice directed flow is indeed the basic mechanism in, for example, the untyped $\pi$-calculus \cite{Milner1999}. Consequently, for this approach to be supported and implementable the langauge does not require to be particularly rich in term of type system and language primitives. Jolie itself originally supported only this mechanism.

\subsection{Data-driven approach}
In this case, the choice operator allows us to use a data-driven approach to computations. 

Let's enrich the interface syntax with the following data type and operation:

\begin{lstlisting}[basicstyle=\small]
//Car rent interface
...
type request: customer | car_return

interface CarRentInterface { 
	...
    RequestResponse: 
      process(request)(string)
}
\end{lstlisting}

\textit{process} operation takes variable of \textit{request} type, which itself can be either of \textit{customer} or \textit{car_return} types 

Let's implement the new server, supporting \textit{process} operation:
\begin{lstlisting}[basicstyle=\small]
include "carRentInterface.iol"

inputPort RentService2 {
	Location: "socket://localhost:2002"
	Protocol: sodep
	Interfaces: CarRentInterface
}

execution{ concurrent }

main{
  process_user_request(request)(response){
  request match {
    customer { response = "43535" };   
    car_return {
      if (request.car_state == "damaged"){
        response = "Car is damaged!"
	  } else {
        response = "Thank you!"
      }
    }
  }
}
\end{lstlisting}

	\textit{Please note that "match" directive is not in the stable version of the project yet, so this example will be not compiled with the current (1.4.1) version of the interpreter.}

In this case we don't need to separate execution of operations by means of processes. Here it is done by means of the type of the input request variable.

In order to test new approach, let's change the client program's code:
\begin{lstlisting}[basicstyle=\small]
//Client.ol
include "carRentInterface.iol"
include "console.iol"

outputPort RentService {
	...
}

outputPort RentService2 {
	Location: "socket://localhost:2002"
	Protocol: sodep
	Interfaces: CarRentInterface
}

main{
  //sending request for a car
  println@Console( "Process-driven approach")();
  ...

  //returing the car
  ...
	
  //working with server based on data type
  println@Console( "Data-driven approach")();
  process@RentService2(customer)(response);
  println@Console
    ( "Car rent request is accepted. ")();
  println@Console 
    ( "Car id is " + response )();
  process@RentService2(return)(response);
  println@Console
   ( "Car is returned. " + response )()
}
\end{lstlisting}

The result of the execution will be the following:
\begin{lstlisting}[basicstyle=\small]
Data-driven approach
Car rent request is accepted. Car id is 43535
Car is returned. Car is damaged!
\end{lstlisting}

The results are clearly the same than the process-driven approach, showing how behaviorally the
two implementations appear undistinguishable for an external observer. However, as we have seen,
the internal computations actually differ, and performances can differ too as long as other 
quality attributes. It is beyond the scope of this paper to make any quantitative assessment.
It is enough to notice how enriched language mechanisms offer alternative programming pattern
-- and therefore desing patterns -- to developers who adopt the service-oriented paradigm. 
The development of specific design guidelines is left as future work.

\section{Architecture of Jolie interpreter}
\label{sec:interpreter}

In order to explain better the changes that were applied to the interpreter, we have to describe how its basic components are working together.

The Jolie interpreter is written in Java, and its architecture is organized into several components \cite{montesi}, the most significant of which regard parsing the source files, establish the communication between components and running them.

The \textit{Parser} part scans and parses the source code, transforms and organizes it as a tree of objects with desired semantics. As a result, parser produces 
\textit{OOIT} (Object oriented interpretation tree), which implements the execution of the semantic rules relative to the input program.
\textit{Runtime environment} instantiates other components and execute the OOIT. And the \textit{Communication core} part is in charge of performing communication between different components abstracting from the communication methods and protocols. 

\begin{figure}[H]
\centering{\includegraphics[scale=0.7]{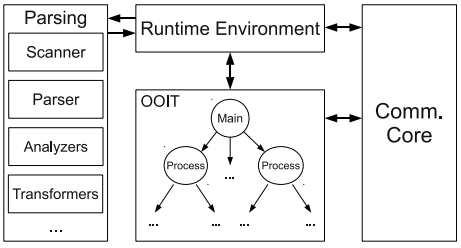}}
\caption{Jolie interpreter architecture}
\end{figure}


This section is dedicated to discussion of interpreter components in more details. 

\subsection{Parser}
The parsing process follows the following stages and involve the following components:
\begin{enumerate}
\item \textit{Scanner} reads input and create token objects based on the ones defined in token types enumeration;
\item \textit{OLParser} is a recursive descent parser, which take created tokens, checks them by the grammar rules and generate the corresponding syntax node in the abstract syntax tree (AST);
\item \textit{OLParseTreeOptimizer} takes ready AST and optimizes it by reducing the number of nodes or transforming the code to more efficient versions; 
\item \textit{SemanticVerifier} checks whether the code is well-formed and semantically correct.
\end{enumerate}


\vspace{0.2cm}
Both OLParserTreeOptimizer and SemanticVerifier use visitor design pattern \cite{Gamma} to access AST (implement OLVisitor class).

Types mentioned in section~\ref{sec:jtp} are expressed as nodes of AST. They are implemented by means of descendants of abstract class TypeDefinition:

\begin{itemize}
\item TypeInlineDefinition for expressing native types;
\item TypeDefinitionLink for linked types;
\item TypeDefinitionUndefined for undefined types.  
\end{itemize}

\subsection{Runtime environment}
OOITBuilder reads the AST and produces object tree-like data structure called object-oriented interpretation tree (OOIT), that defines the semantics for program execution. OOIT nodes implement Process interface, so that each node is responsible for the semantics of a single statement.


\vspace{0.2cm}
Runtime Environment handles parallel execution by means of native threads. Threads responsible for executing a part of the OOIT can be of two types: 
\begin{enumerate}
\item Session threads, which are used for handling different sessions and retain a local state for variable values.
\item Parallel threads, which are used for handling parallel composition and refer to their parent session thread for state handling.
\end{enumerate}

\subsection{Communication core}
Communication Core component is used for performing communications by means of such mechanisms as messages and channels. Channels are used for the sending and receiving of messages, which consist of resource path, name of operation they are dedicated for, message content and in some cases fault name. Channels are in charge of encoding/decoding messages using the right protocol and sending/receiving them by means of the right communication medium.


\section{Towards data-driven Workflows for Microservices}
\label{sec:data-driven}
Jolie still experiences lack of some data types and operations, which could enrich its syntax and add extra flexibility, like, for example, regular expressions or choice operator. This work is dedicated to implementing the choice operator as the one which is proved to be useful.

The idea of choice operator was taken from XML[7]. Choice operator in XML allows to put several elements in choice element declaration, but to be presented with only one of them. For example, there is a choice element called "animal" in example below, which can be either presented as a dog or as a cat:

\begin{lstlisting}[basicstyle=\small]
<xs:element name="animal">
  <xs:complexType>
    <xs:choice>
      <xs:element name="cat" type="cat"/>
      <xs:element name="dog" type="dog"/>
    </xs:choice>
  </xs:complexType>
</xs:element>
\end{lstlisting}

Despite choice element in XML allows to choose between several types, in our implementation, this construction provides possibility of choosing between two types only.
The same example will look in Jolie language in the following way:

\begin{lstlisting}[basicstyle=\small]
type animal: cat | dog
\end{lstlisting}

We use pipe character ("|") for choice operator. 
Note, that in this case cat and dog are linked types and need to be declared explicitly or otherwise Jolie interpreter will raise an exception.

TypeDefinition formal grammar was enriched correspondingly:   
\vspace{0.2cm}

{\setlength\tabcolsep{4pt}
\begin{tabular}{c >{$}l<{$} l l}
TypeDefinition &\Coloneqq & $\ldots$ \\
   & | & TypeDefinition | TypeDefinition
\end{tabular}}
\vspace{0.2cm}

\subsection{Architectural changes}
In this section we will describe the architectural changes necessary to implement the new data type.

\subsubsection{Adding choice type to AST}
In order to support possibility of storing two types, TypeChoiceDefinition class was created (extends TypeDefinition as TypeInlineDefinition and others).

\vspace{0.2cm}
\begin{figure}[H]
\centering{\includegraphics[scale=0.50]{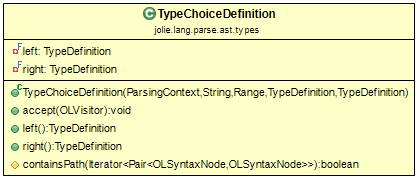}}
\caption{TypeChoiceDefinition class diagram}
\end{figure}

\vspace{0.2cm}
The TypeChoiceDefinition class contains two attributes left and right representing two possible types, that can be any of TypeDefinition members. The addition of this class requires a change in the parent TypeDefinition class and its other descendants, as they contain several methods, that are not applicable to choice type (like checking if a type has subtypes or its native type), and which might raise null reference exception while trying to access them. This is why these abstract methods were removed.

\vspace{0.2cm}
\subsubsection{Extending parser}
Extension of parser requires working on Scanner, OLParser, OLParserTreeOptimizer and SemanticVerifier. 

\textit{Scanner.} No changes to the scanner are needed, because the pipe-symbol ("|"), which represents choice operator, is already in use as the parallel operator, so it is already present in the  TokenType enum. Parallel operator, placed between operands, indicates that they are are executed concurrently.

\begin{lstlisting}[basicstyle=\small]
statementA | {statementB; statementC}
\end{lstlisting}

However, execution relates to the behaviour part of the program, while message types have to be defined in the deployment part, and they are separately processed, so no effort on pipe symbol processing redefinition is needed. 

\textit{OLParser} parses message types and their subtypes by means of parseTypes and parseSubTypes methods correspondingly. These methods check that the sequence of tokens, generated by Scanner, is correct regarding the type definition grammar, as presented in section~\ref{sec:jl} and then generate a node in AST of one of the TypeDefinition descendants. Since this grammar has been enriched, parseTypes and parseSubTypes methods have been changed correspondingly, so that they could expect "|" token inside the type definition.

As \textit{OLParserTreeOptimizer} and \textit{SemanticVerifier} classes works with AST by means of visitors, so visit method for TypeChoiceDefinition type was added to OLVisitor class. In this case, no specific optimization is needed, so this visitor simply does the same actions as the other TypeDefinition class descendants. 

\textit{SemanticVerifier} checks whether the types with the same name have been already defined and checks  cardinality of types. New visit method, added to deal with objects of TypeChoiceDefinition type, is doing the same, except that it also invoke semantic verification on both of its types inside.

\vspace{0.2cm}
\subsubsection{Extending runtime facilities}
Extending runtime facilities regards mostly building OOIT process and message types themseleves.

\textit{Additions to building object-oriented interpretation tree process.} OOITBuilder generates object oriented interpretation tree, based on AST produced by OLParser. It also uses visitors to access nodes of AST. Since new TypeChoiceDefinition type was added, corresponding visitor was created, which is in charge of creating type object based on current TypeChoiceDefinition object. Type objects are discussed in the next section.

\textit{Message types.} As messages are being passed by means of type objects, each of the TypeDefinition descendants in AST should have corresponding representation in Type class descendants (TypeImpl for TypeInlineDefinition, TypeLink for TypeDefinitionLink).

For representing TypeChoiceDefinition, new Type descendant, TypeChoice, has been created.

\vspace{0.2cm}
\begin{figure}
\centering{\includegraphics[scale=0.50]{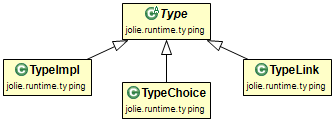}}
\caption{Jolie type system class diagram}
\end{figure}

\subsection{Examples}
\label{sec:ex}
\subsubsection{Working with types and subtypes}
Choice operator can work with native types:

\begin{lstlisting}[basicstyle=\small]
type numeric: int | long
\end{lstlisting}

Linked types:
\begin{lstlisting}[basicstyle=\small]
type linked_type: string
type linked_choice: linked_type | void
\end{lstlisting}

Subtypes:
\begin{lstlisting}[basicstyle=\small]
person_info: void {.id: string | int}
\end{lstlisting}

\subsubsection{Functions genericity}
One of the way of choice operator utilization is for making functions generic. If we declare the following choice type and function in interface 

\begin{lstlisting}[basicstyle=\small]
type choice: string | int
fun_choice(choice)(choice)
\end{lstlisting}

We can provide the same behaviour without considering which type (string or int)  was passed to the function:

\begin{lstlisting}[basicstyle=\small]
fun_choice(request)(response){
	response = request
}
\end{lstlisting}

Or imagine that we need to implement different behaviour of the function based on the arguments passed. 
Let's have an interface with type person, able to be presented as element with type personSSN or personCCN:

\begin{lstlisting}[basicstyle=\small]
type person: personSSN | personCCN
type personSSN:void {
	.ssn:int
}
type personCCN:void {
	.ccn:string
}
\end{lstlisting}

And a function pay, which takes input argument of person type and run the corresponding code based on particular type (personSSN or personCCN) of the argument passed.

\begin{lstlisting}[basicstyle=\small]
pay(person)(response) {
	person match {
	  personCCN: ...
	  |
	  personSSN: ...
	}
	if ( is_defined( person.ssn ) ) {
	  // ask the person registry
	} else {
	  // contact the bank
	}
}
\end{lstlisting}

It is also possible to use choice operator if you need two support several versions of data structure. For example, it was needed to process data related to Old-Software corporation, later its title has been changed to New-Software, but some customers can still use the old one.  

\begin{lstlisting}[basicstyle=\small]
type Old-Software-Corp: void {.
name: string
address: string
}

type New-Software-Corp: void {.
name: void {
.firstname: string
.lastname: string
address: string
phone: int
}

type corporation: Old-Software-Corp 
	| New-Software-Corp
\end{lstlisting}

\section{Conclusions and Future work}
\label{sec:conclusions}
Jolie is a comprehensive programming language based on the service-oriented paradigm~\cite{jolie}, which emerged in the context of an extended research effort aimed at formalizing Service-Oriented Computing on top of broadly accepted models of concurrency in the EU Project SENSORIA (see, e.g., \cite{Guidi07,LucchiM07}). Due to its support for the quick prototyping of both simple services and complex service coordination, Jolie has been used in the development of other research projects involving the programming and deployment of services (including~\cite{appliedchor,aiocj}).
However, it has been identified how the language still lacks of data types and operations able to enrich the syntax and add extra flexibility to program common SOA scenarios. Regular expressions and choice operator are just examples of this deficiency. This work has been devoted to extend the Jolie type system in order to add the choice operator and realize the necessary changes into the interpreter.

The major outcomes can be summarized as follows:

\begin{itemize}
\item Identification of data types able to support common SOA programming scenarios
\item Addition of choice operator in the syntax and semantics of the language
\item Analysis and reengineering of Jolie interpreter
\end{itemize}

This work shows how computation can move from a process-driven to a data-driven approach. Control-flow can be now directed at the time of message passing and according to the nature of the message structure and type, instead of requiring post-reception processing, hence leading to a preliminary identification of recurring communication patterns that can be then shaped as design patterns. In the microservice scenario this represent a novel opportunity opening to new programming scenario.

Future work leaves space to both theoretical investigation and practical realization. Implementing regular expressions is a natural step in order to further enrich the type system and manage a broader set of programming scenarios. For what concerns theoretical aspects, formalization of the extended type system is considered a priority.

\section*{Acknowledgements}
We would like to thank Innopolis University for logistic and financial support. This work was also partially supported by CRC (Choreographies for Reliable and efficient Communication software), grant no. DFF--4005-00304 from the Danish Council for 
Independent Research. Our gratitude goes to colleagues of the Institute of Technologies and Software Development who participated in the discussion and the seminar, in particular Bertrand Meyer, Victor Rivera, Daniel de Carvalho, Mohamed Elwakil, Leonard Johard, Alexander Naumchev, Alexander Chichigin and Rasul Tumyrkin.

\end{document}